\documentclass[%
 reprint,
superscriptaddress,
 amsmath,amssymb,
 prl,
]{revtex4-1}

\usepackage[dvipsnames]{xcolor}

\usepackage{graphicx}
\usepackage{dcolumn}
\usepackage{bm}

\begin{document}


\title{
Relating Chain Conformations to Extensional Stress In Entangled Polymer Melts
}

\author{Thomas C. O'Connor}
 \email{toconnor@jhu.edu}
 \affiliation{Department of Physics and Astronomy, Johns Hopkins University}
 
\author{Nicolas J. Alvarez}%
 \email{nja49@drexel.edu}
\affiliation{Department of Chemical and Biological Engineering, Drexel University}

\author{Mark O. Robbins}
 \email{mr@jhu.edu}
 \affiliation{Department of Physics and Astronomy, Johns Hopkins University}

\date{\today}

\begin{abstract}
Nonlinear extensional flows are common in polymer processing but remain challenging theoretically because dramatic stretching of chains deforms
the entanglement network far from equilibrium. 
Here, we present coarse-grained simulations of extensional flows in entangled polymer melts for Rouse-Weissenberg numbers $Wi_R=0.06$--$52$ and Hencky strains $\epsilon\geq6$.
Simulations reproduce experimental trends in extensional viscosity with time,
rate and molecular weight.
Studies of molecular structure reveal an elongation and thinning of the confining tube with increasing $Wi_R$.
The rising stress is quantitatively consistent with the decreasing entropy of chains at the equilibrium entanglement length.
Molecular weight dependent trends in viscosity are
related to a crossover from the Newtonian limit
to a high rate limit that scales differently with chain length.
\end{abstract}

\pacs{Valid PACS appear here}
\maketitle
The development of an accurate molecular model for polymer dynamics in complex flows has been the focus of intense research for more than 50 years.
The tube model \cite{Doi1988} has been incredibly successful in describing the dynamics of polymers in terms of entanglements with other chains that form a temporary confining tube that hinders diffusion.
It captures the linear response of chemically distinct melts in terms of a universal function of the number of entanglements per chain $Z$
and a material specific entanglement time $\tau_e$ and stress scale $G_e$ \cite{Likhtman2002}.
However a growing number of experiments show new physics must be incorporated to understand strongly nonlinear flows that are relevant to industrial applications.
A striking example is that melts with identical $Z$ and linear response can show
opposite trends in strong elongational flows, with viscosity rising or falling with increasing rate
\cite{Bach2003,Nielsen2006,Huang2013a,Huang2013b,Wingstrand2015}.

A number of attempts have been made to generalize the tube model based on different hypotheses about molecular mechanisms, including convective constraint release under shear \cite{Marrucci1996}, segmental stretch \cite{Mead1995,Mead1998},
interchain pressure \cite{Marrucci2004}, formation and destruction of ``slip-links'' \cite{Likhtman2005, Baig2010,Andreev2013,Ianniruberto2014b}, and friction reduction in elongational flows\cite{Ianniruberto2012,Yaoita2012,Wingstrand2015,Costanzo2016}.
To date, no generalization of the tube model has been able to predict behavior in strong elongational flows
\cite{Bach2003,Huang2013a,Huang2013b,Wingstrand2015}
and experiments have not provided direct measures of changes in chain conformation.
It remains unclear how the confining tube changes in nonlinear flows,
how chain conformations affect
dissipation and what role chemistry \cite{Sridhar2013}, $Z$, chain length, and the equilibrium entanglement length $N_e$ play in determining the molecular mechanisms underlying nonlinear behavior.

Molecular dynamics (MD) simulations are an ideal platform for relating macroscopic response to molecular structure \cite{Likhtman2007,Cao2015}, but it has been difficult to simulate strong elongational flows at sufficiently large strains
to reach steady state \cite{Xu2018, Kroger1997,Daivis2003}.
In this paper we use a recently developed technique \cite{Dobson2014,Nicholson2016} to overcome this barrier.
The simulations capture experimental trends in both the transient and steady state nonlinear viscosity of melts with different $Z$ and entanglement length $N_e$ \cite{Sridhar2013,Wingstrand2015}.
Trends in viscosity with rate, $Z$ and $N_e$ are explained as
a cross-over from the Newtonian limit to a high rate limit for aligned chains.
A simple scaling law for the high rate behavior is derived and verified.
The observed macroscopic response is shown to arise from changes in chain statistics that can be described as alignment and contraction of a confining tube with increasing strain rate.
Surprisingly, changes in segment orientation only depend on the degree of entanglement $Z$,
while chain stretching at high rates only depends upon the equilibrium entanglement length $N_e$.
For all rates and melts,
the steady-state stress is quantitatively related to changes in chain entropy over segments of length $N_e$.

%
Polymers are modeled with the well-studied Kremer-Grest bead-spring model \cite{Kremer1990} using LAMMPS \cite{Plimpton1995}.
All beads interact with a truncated Lennard-Jones (LJ) potential and results are presented in reduced LJ units.
Linear chains of $N$ beads are bound together with a FENE potential with mean bond length $b \approx 0.96$.
To vary tube model parameters, the chain stiffness is controlled by a bond bending potential
$k_{bend} (1- \cos \alpha)$, where $\alpha$ is the angle between successive bonds.
For the melt labeled M1, $k_{bend}=1.5$
and there are $N_e\approx28$ beads per rheological entanglement length \cite{Moreira2015}.
For melt M2, $k_{bend}=0.75$ and $N_e\approx51$
\cite{Moreira2015}.

Melts with $M$ chains
are equilibrated at temperature $T=1$ and density $\rho=0.85$ with standard methods \cite{Auhl2003}.
M1 melts have $M=1640$, $N=112$; $M=1094$, $N=168$; $M=734$, $N=250$; or $M=368$, $N=500$, corresponding to $Z \approx 4$, $6$, $9$ or $18$, respectively.
M2 melts have M=918, N=200; M=354, N=300 or M=405, N=450, corresponding to $Z\approx4$, 6 or 9, respectively.
Melts are deformed
at constant Hencky strain rate $\dot\epsilon \equiv \partial\ln\Lambda/\partial t$ with $\Lambda$ the stretch along the z-axis.
Since polymers
are nearly incompressible, the two perpendicular directions contract by $1/\sqrt{\Lambda}$.
Flow is maintained by integrating the SLLOD equations of motion, and Generalized Kraynik-Reinelt boundary conditions are used to prevent the simulation box from becoming too small in the perpendicular directions \cite{Dobson2014,Nicholson2016}.
As shown in Supplemental Material (SM) Fig. S1, simulations reproduce the time-dependent evolution of viscosity that is observed in experiments, while achieving strains that are not currently experimentally accessible.
The steady-state extensional stress
$\sigma_{ex}=\sigma_{zz}-\frac{1}{2}(\sigma_{xx}+\sigma_{yy})$
and chain statistics are obtained by averaging 
simulation data over the strain interval $\epsilon \in [5.5,6.0]$.
 
Experiments typically plot dynamic viscosity data in reduced units based on tube theory to facilitate comparison between different melts \cite{Dealy2006,Sridhar2013,Wingstrand2015}.
Times are scaled by the entanglement time $\tau_e$ and the viscosity by $G_e \tau_R$, where the Rouse time $\tau_R=\tau_e Z^2$ is the characteristic time for a stretched chain to relax to its equilibrium contour length and $G_e=\rho k_B T/N_e$ is the entanglement modulus.
A dimensionless measure of flow rate is the Rouse-Weissenberg number $Wi_R=\dot\epsilon\tau_R$. 
Previous studies have measured $\tau_e\approx1.98\times 10^3$ and $6\times 10^3$ for M1 and M2 melts, respectively \cite{Moreira2015,Hsu2016,Ge2014}.

\begin{figure}
    \centering
    \includegraphics[width=\linewidth]{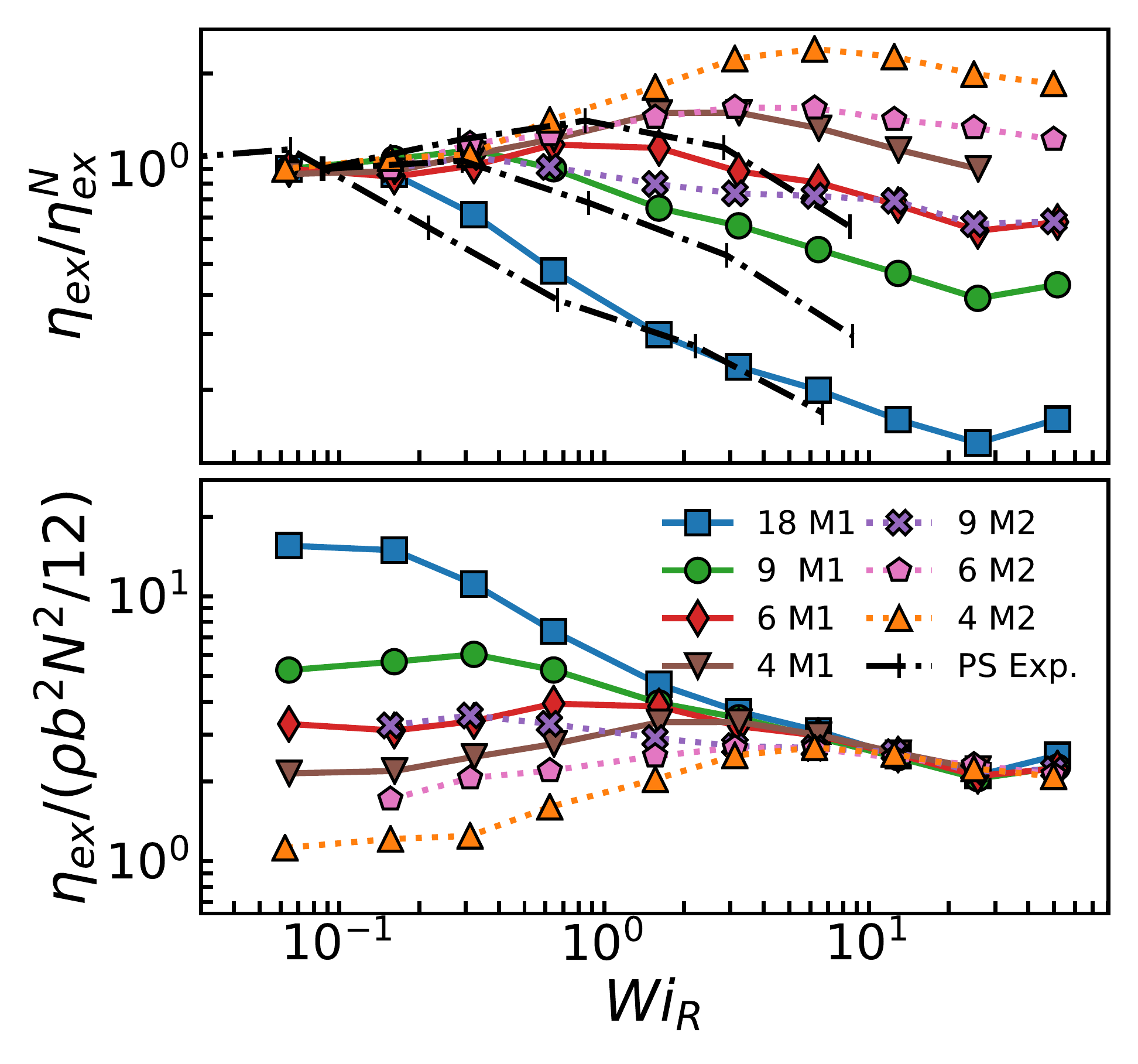}
    \caption{(a) Ratio of steady state viscosity $\eta_{ex}$ to Newtonian viscosity vs. $Wi_R$ from simulations of M1 and M2 melts at the indicated $Z$ and experiments on PS
    at $Z=7.5$ \cite{Nielsen2006}, $10$ \cite{Costanzo2016}, and $21$ \cite{Wingstrand2015} (top to bottom).
    (b) Simulation data from panel (a) renormalized by the formula for the asymptotic drag on straight chains. Results for all chains collapse at large $Wi_R$.
    \label{fig:steady}
    }
\end{figure}

Figure \ref{fig:steady}(a) shows the rate dependence of the steady state viscosity
$\eta_{ex}\equiv \sigma_{ex}/\dot\epsilon$ normalized by the
value in the Newtonian limit $\eta^N_{ex}$.
Data for M1 and M2 melts are shown alongside experimental results for polystyrene (PS) at similar $Z$.
The significant difference between results for M1 and M2 melts at any common value of $Z$ is consistent with the deviations from tube theory found in past experiments \cite{Wingstrand2015}.
Although $Z$ is not enough to determine the nonlinear response,
all melts show common trends with increasing $Z$.
For both simulations and experiments,
the longest chains begin to shear-thin at the lowest $Wi_R$ and show the largest drop in viscosity.
The decrease is almost an order of magnitude for M1 and PS at $Z=18$ and 21, respectively.
As $Z$ decreases, the onset of shear thinning moves to larger $Wi_R$ and the decrease in $\eta_{ex}$ decreases.
Indeed shorter chains show some initial shear thickening for all systems \cite{Nielsen2006,Wingstrand2015,Costanzo2016}.

\begin{figure}
    \centering
    \includegraphics[width=\linewidth]{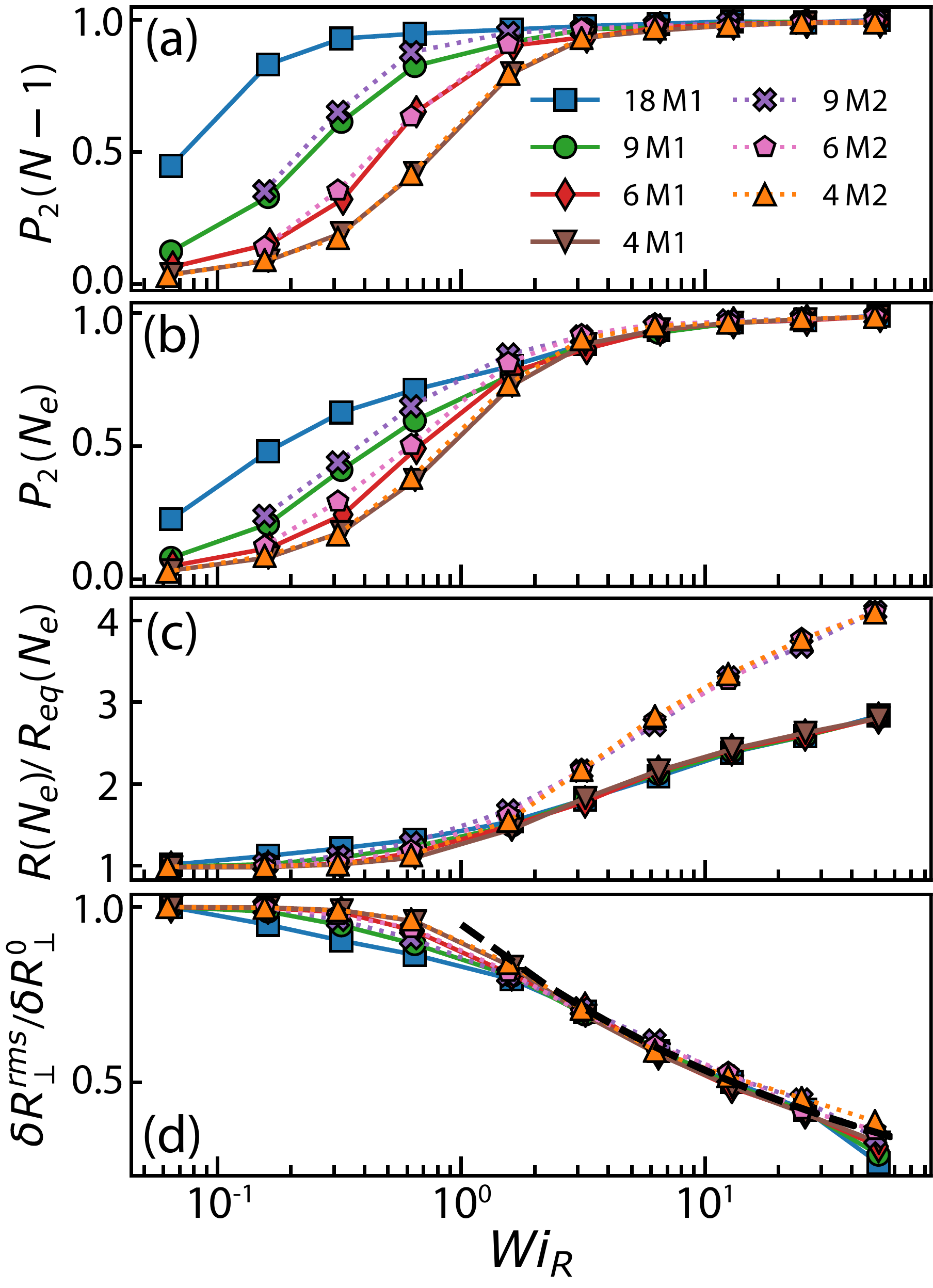}
    \caption{
        (a) Orientational order at the end-end scale $P_2(N-1)$ vs. $Wi_R$ for M1 and M2 melts at the values of Z indicated in the legend.
    (b) Orientational order at the scale of equilibrium entanglements.
    (c) Measure of tube elongation given by stretch of tube segments of length $N_e$.
(d) Measure of decrease in tube radius given by change in rms
deviation of monomers from the line between endpoints of segments of length $N_e$.
A black dashed line indicates $\sim Wi_R^{1/4}$ scaling.}
   \label{fig:orient}
\end{figure}

Simulations allow us to directly correlate these changes in macroscopic response with changes in molecular structure.
Snapshots in Fig. S2 show chains evolve from nearly equilibrium random coils at $Wi_R << 1$
to nearly straight configurations by $Wi_R=51.5$.
This
evolving orientational order can be described by the nematic order parameter $P_{2}(n)=\frac{1}{2}\left< 3\langle\cos^2 \theta_n\rangle-1 \right>$,
where $\left< \right>$ indicates an ensemble average and $\theta_n$ is the angle between the extension axis and the vector $\vec{R}(n)$ between beads separated by $n$ bonds.
As shown in Fig. \ref{fig:orient}, $P_2(n) =0$ for randomly
oriented chains at low $Wi_R$ and approaches unity at high $Wi_R$, corresponding to complete alignment.
Alignment occurs first at the full chain length ($N-1$ bonds) and affects smaller $n$ as $Wi_R$ increases. 
The rate where $P_2(N-1)$ approaches unity coincides with the onset of a decrease
in $\eta_{ex}/\eta_N$.
Both changes imply that deformation is faster than the the longest relaxation time, the disentanglement time
$\tau_d$ for chains to escape their tubes.\cite{Doi1983,Hou2010}
As shown in Fig. S3, $P_2(N-1)\approx 0.3$ for $\dot\epsilon \tau_d =1$ and saturates near unity for $\dot\epsilon \tau_d \gtrsim 5$.

Alignment at the entanglement scale starts at larger $Wi_R$.
In the low rate regime ($Wi_R \leq 1$), $P_2(N_e)$ is the same for different melts at the same $Z$.
By $Wi_R=1$, $P_2(N_e)$ exceeds 0.5 for all melts, and the results collapse on to a common curve  for $Wi_R >> 1$.
For $\dot\epsilon \tau_d > 1$ chains are deformed faster than they can escape their tubes.
For $Wi_R>1$ the tube is being deformed and aligned along the extension direction faster than chains can relax to their equilibrium length along the tube.
As a result, segments are stretched and aligned at shorter and shorter scales as $Wi_R$ increases.

\begin{figure}
    \centering
    \includegraphics[width=\linewidth]{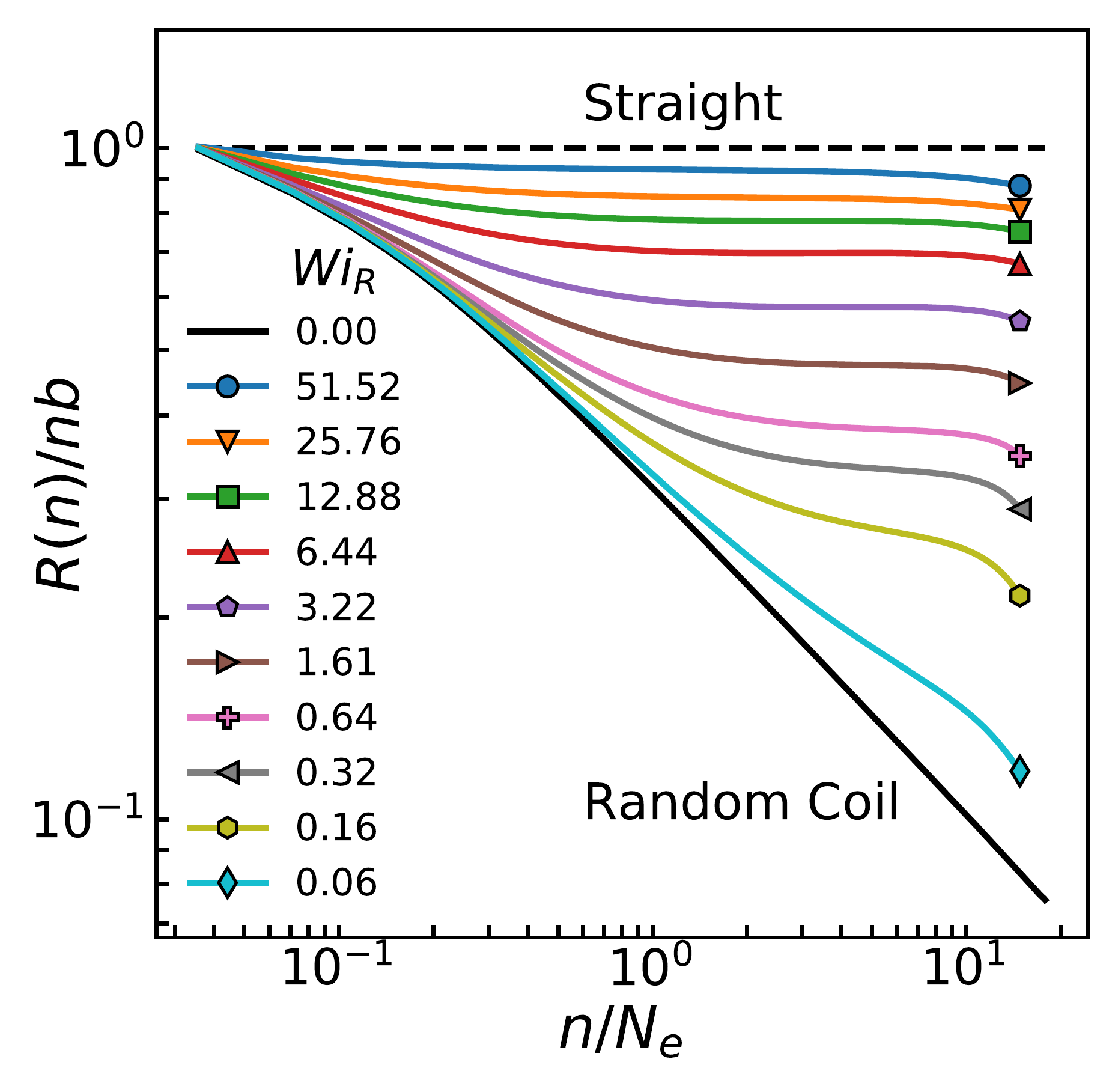}
    \caption{
    Ratio of rms length to contour length $R(n)/nb$ as a function of $n$ for a $Z=18$ M1 melt at the indicated $Wi_R$. The solid black line corresponds to the equilibrium coil structure with $R(n)/n\sim (C_\infty/n)^{1/2}$ for $n$ larger than $C_\infty$. Flow straightens chains at scales larger than $N_e$ for $Wi_R<1$. For $Wi_R>1$, the chain becomes straight on progressively smaller scales. 
The small drop at large $n$ indicates that chains are less aligned within an entanglement or two from their ends.
\label{fig:internal}
}
\end{figure}

A measure of straightening is provided by $R(n)$ the root mean squared (rms) length of $\vec{R}(n)$.
This must be less than the contour length between beads $nb$,
where the bond length $b$ remains essentially
unchanged at the highest $Wi_R$ considered here.
The fraction of the fully extended length $R(n)/nb$ is shown in Fig. \ref{fig:internal}.
At $Wi_R=0.06$ chains have a near equilibrium conformation.
The ratio $R_{eq}(n)/nb$ decreases slowly with increasing $n$ at small $n$
because $k_{bend}$ makes the chain fairly straight.
The behavior changes above the Kuhn length $\ell_K \equiv b C_\infty$, where
the chain stiffness constant $C_\infty$ is 2.8 and 2.2 for M1 and M2, respectively \cite{Hsu2016}.
At larger $n$, chains are random coils and $R_{eq} (n)/nb= \sqrt{ C_\infty/n}$
(black solid line in Fig. \ref{fig:internal}).
For $Wi_R = 0.06$ short chains follow the equilibrium behavior at all $n$.
For the longer chains shown, there is a slight straightening and reduced rate of decrease in $R(n)/nb$ at large $n$ because $\dot \epsilon \tau_d \sim 1$.

The scale at which there is significant straightening decreases from $N$ to $N_e$ as $Wi_R$ increases towards unity.
For $Wi_R>1$ the behavior is qualitatively different, with $R(n)/nb$ saturating at large $n$.
Saturation starts near $N_e$ for $Wi_R=1.61$ and moves to smaller $n$ as
$Wi_R$ increases.
These results are exactly as expected from the snapshots in Fig. S2.
Flow increases the length of chains and reduces fluctuations around their
end-end vector.

The changes in $R(n)$ imply changes in the conformation of the tube confining each chain.
In equilibrium the tube has a radius of order $R_{eq}(N_e)$ and is a random walk at larger scales with a Kuhn length of order $R_{eq}(N_e)$.
Under elongation, Fig. \ref{fig:orient} shows that the tube stretches and narrows. 
The increase in tube length per $N_e$ can be measured by calculating the ratio
$R(N_e)/R_{eq}(N_e)$.
As shown in Fig. \ref{fig:orient}c, the tube length remains nearly constant as $Wi_R$ increases to unity and then rises rapidly.
Results for each melt collapse on to a common curve and at large $Wi_R$ approach the maximum possible stretch $\lambda_{max}=N_e b/R_{eq}(N_e)=3.16$ and 4.8 for M1 and M2, respectively.

To characterize the change in tube diameter we consider segments of length $N_e$ and evaluate the maximum rms fluctuation $\delta R_\perp$ in the plane perpendicular to the end-end vector.
Only the central 20\% of each segment is included because the
	fluctuation goes to zero at the ends of each segment (see SM Sec. D).
As for the tube length, the tube radius begins to change rapidly for $Wi_R >1$ (although there is a small change for the longest chains when $\dot\epsilon \tau_d >1$).
The fractional change in tube radius at $Wi_R >1$ is nearly the same for all $Z$ and both melts. 
In all cases $\delta R_\perp$ is of order $b$ at the largest $Wi_R$, corresponding to a nearly straight chain.

Short segments are able to retain a nearly equilibrium conformation up to higher $Wi_R$.
Since the relaxation time scales as $n^2$, one may expect that the relaxed
length $n_{rel}(Wi_R)$ scales as $Wi_R^{-1/2}$.
The length of the corresponding random walk should set the tube radius,
leading to a prediction that $\delta R_\perp \propto Wi_R^{-1/4}$ until
the radius approaches the bond length.
The tube length should grow as the number of segments of length $n_{rel}$ times the rms length of each, implying $R(N_e)/R_{eq}(N) \propto (N_e/n_{rel}) (n_{rel}/N_e)^{1/2} \propto N_e^{1/2} Wi_R^{1/4}$ until it saturates at $\lambda_{max}$.
The scaling range is not large enough to accurately test these scaling exponents, but the thick dashed line in Fig. \ref{fig:orient}(d) shows that the decrease in $\delta R_\perp$ is consistent with a $-1/4$ power law \cite{Marrucci2004}
\footnote{Marrucci and Ianniruberto proposed a theory predicting $\delta R_\perp \propto Wi_R^{-1/4}$ based on an inter-chain tube pressure. However, the dynamics of their theory predicts a cross-over to this scaling controlled by $\tau_d$. This cannot describe our data which shows $\delta R_\perp$ only depends on $\tau_R$ for $Wi_R>1$.}.

The alignment of chains by flow implies a reduction in their entropy that contributes to the steady-state stress $\sigma_{ex}$.
While there is also an energetic contribution, it is much smaller because there is almost no stretching of chain backbones at the highest $Wi_R$ considered here.
One can calculate the stress due to the entropic force $\vec{F}(n)$ of segments of length $n$ stretched to $\vec{R}(n)$. 
The density of such segments is $\rho/n$,
where $\rho$ is the monomer density.
The stress tensor
$   \sigma_{ij}=(\rho/n) \left<          F_i(n)R_j(n) \right>
$
where $i$ and $j$ are cartesian coordinates \cite{Rubinstein2003}.
Since the force is directed radially, the extensional stress is
$   \sigma_{ex}=(\rho/n) \left< R(n) F(n) P_2(n) \right>
$
Inserting the standard result for the force on a random chain yields
\begin{equation}
    \sigma^{ent}_{ex}(n) = \frac{\rho k_B T}{C_\infty} \left< \frac{R(n)}{nb} L^{-1}\left( \frac{R(n)}{nb}\right) P_2(n) \right>
    \label{eq:predict}
\end{equation}
where the inverse Langevin function $L^{-1}$ accounts for the nonlinear reduction in entropy as segments approach full extension \cite{Doi1988,Rubinstein2003}.

In the Newtonian limit, the tube model relates the stress to the change in entropy of segments with length $n \sim N_e$.
A network of entanglements is assumed to carry the stress at larger scales \cite{Doi1988}.
It is not clear whether the same $n$ should be used in Eq. \ref{eq:predict} for the highly aligned states at large $Wi_R$.
However, $\sigma_{ex}^{ent}(n)$ is insensitive to $n$ at large $Wi_R$ because of the plateau in $R(n)/nb$ (Fig. \ref{fig:internal}).
Figure 4 plots the total measured steady-state stress versus $\sigma^{ent}_{ex}(N_e)$ for all simulated liquids in steady state.
The two quantities are in excellent agreement for all melts and over three orders of magnitude in reduced stress.
Deviations only become significant at the largest $Wi_R$ where the chains are
nearing complete alignment and Eq. \ref{eq:predict} becomes singular.

\begin{figure}
    \centering
    \includegraphics[width=\linewidth]{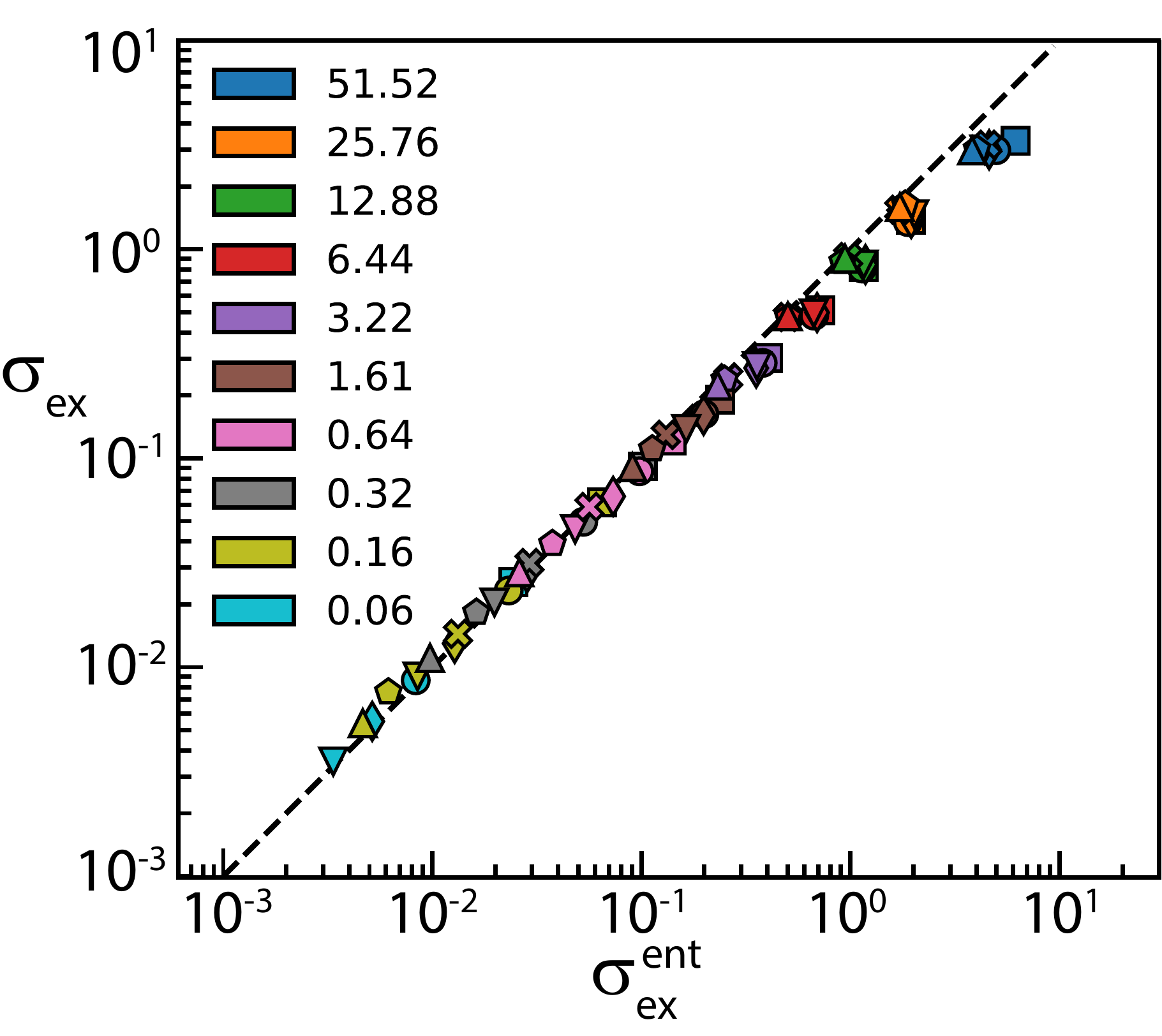}
    \caption{Comparison of steady state stress from simulations $\sigma_{ex}$ to the entropic stress from Eq. \ref{eq:predict}. Colors correspond to the values of $Wi_R$ in the legend and symbols indicate the melt.
    M1 at $Z=18$ (squares), $9$ (circles), $6$ (diamonds), and $4$ (down triangles). M2 at $Z=9$ (X), $6$ (pentagons), $4$ (up triangles).}
    \label{fig:entropy}
\end{figure}

To obtain the viscosity in the large $Wi_R$ limit we equate the macroscopic
rate of dissipation per unit volume $\eta_{ex} \dot{\epsilon}^2$ to the
microscopic dissipation.
Since chains are nearly fully extended, we consider a single straight chain in an extensional uniaxial flow.
The entire chain must have the same average velocity, so the mean velocity $\Delta v$ of its monomers relative to their neighbors grows linearly with distance $x$ from the chain center as $\Delta v=\dot\epsilon x$.
If there is a linear drag force with drag coefficient $\zeta$,
each monomer dissipates energy at a rate $\zeta \Delta v^2$.
Averaging the dissipation over $x$ gives a dissipation per monomer of
$\left< \zeta \Delta v ^2\right> = 
\zeta b^2 N^2 \dot{\epsilon}^2/12$
and thus $\eta_{ex}=\rho \zeta b^2 N^2/12$.
In general, $\zeta$ will depend upon the chemical structure and interactions of the chain backbone.

Figure \ref{fig:steady}(b) shows $\eta_{ex}$ normalized by $\rho b^2 N^2/12$.
Results for both models and all chain lengths collapse onto a universal curve 
at large $Wi_R$ whose limiting value corresponds to $\zeta$.
Note that M1 and M2 are expected to have nearly the same $\zeta$ because they have identical monomer masses, bond lengths and interchain interactions.
The main difference is that M2 is more rigid and this becomes irrelevant for aligned chains.
The derived value of $\zeta \approx 2$ is consistent with the viscosity of short chains.

The above results explain many experimental observations on 
the nonlinear response of polymers under strong elongational flow and relate them to changes in chain conformation.
At all $Wi_R$ the stress is quantitatively described by the entropic forces associated with chain straightening on segments of length $N_e$.
This entropic stress is balanced by drag forces that also depend on chain conformation and scale with different powers of $Z$ at low and high $Wi_R$.

For small $Wi_R$, chains remain close to Gaussian random walks.
As predicted by the tube model and shown in Fig. S6 \cite{Doi1983}, $\eta_{ex}\sim \eta_N$ and scales as $G_e \tau_e Z^x \propto \zeta_N N_e Z^x$ where $x\approx3.4$ for well entangled chains and $\zeta_N$ is the monomer drag in the Newtonian limit.
At high rates, chains are straight and $\eta_{ex}$ rises as $\zeta N^2$.
The ratio between the Newtonian and high-rate viscosity scales as $\sim Z^{1.4}\frac{\zeta_N}{N_e\zeta}$.
The $Z$ dependence explains why the amount of extension rate thinning increases with chain length in both experiments \cite{Bach2003,Nielsen2006,Wingstrand2015}
and simulations (Fig. \ref{fig:steady}).
The thinning of long chains can be fit to a power law with a $Z$ dependent exponent over about one decade in rate.

Chemistry enters through $N_e$ and the drag coefficients.
Rate thickening may be observed at small $Z$ for melts like M2 with small $\frac{\zeta_N}{N_e \zeta}$.
The increased stiffness of M1 chains, decreases $N_e$ and increases the amount
of thinning.
Diluting with short chains increases $N_e$, which may be part of the reason
solutions show less shear thinning\cite{Ianniruberto2012,Yaoita2012,Wingstrand2015,Costanzo2016}.
There is also evidence that solutions suppress changes in drag because
solvent molecules are less aligned by flow \cite{Ianniruberto2012,Yaoita2012,Wingstrand2015,Costanzo2016}.
Changes in monomer drag with alignment as $Wi_R$ increases are
small in our systems because the monomers are spherical.
Large effects may be expected for polymers with large and rigid side groups.

Many recent methods identify entanglements with contacts between the primitive paths of polymers \cite{Everaers2004,Kroger2005,Tzoumanekas2006}.
These methods suggest that there are no entanglements between the highly aligned chains at large $Wi_R$
\cite{Likhtman2005, Baig2010,Andreev2013,Ianniruberto2014b}.
However, our measured chain statistics show that chains are highly confined at large $Wi_R$ and the volume of the tube associated with the length and radius in Fig. \ref{fig:orient} decreases as $Wi_R$ rises.
Studies of chain relaxation will play an important role in unraveling the relationship between entanglements and the confining tube at high rate.

\begin{acknowledgements}
The authors would like to thank Peter D. Olmsted for valuable discussions. This work was performed within the Center for Materials in Extreme Dynamic Environements (CMEDE) at the Hopkins Extreme Materials Institute with financial Support provided by grant No. W911NF-12-2-0022.
\end{acknowledgements}

\newcommand{\beginsupplement}{%
        \setcounter{table}{0}
        \renewcommand{\thetable}{S\arabic{table}}%
        \setcounter{figure}{0}
        \renewcommand{\thefigure}{S\arabic{figure}}%
     }

\beginsupplement
\onecolumngrid
\section*{SA. Transient Viscosity}
\begin{figure}[h]
    \centering
    \includegraphics[width=0.5\textwidth ]{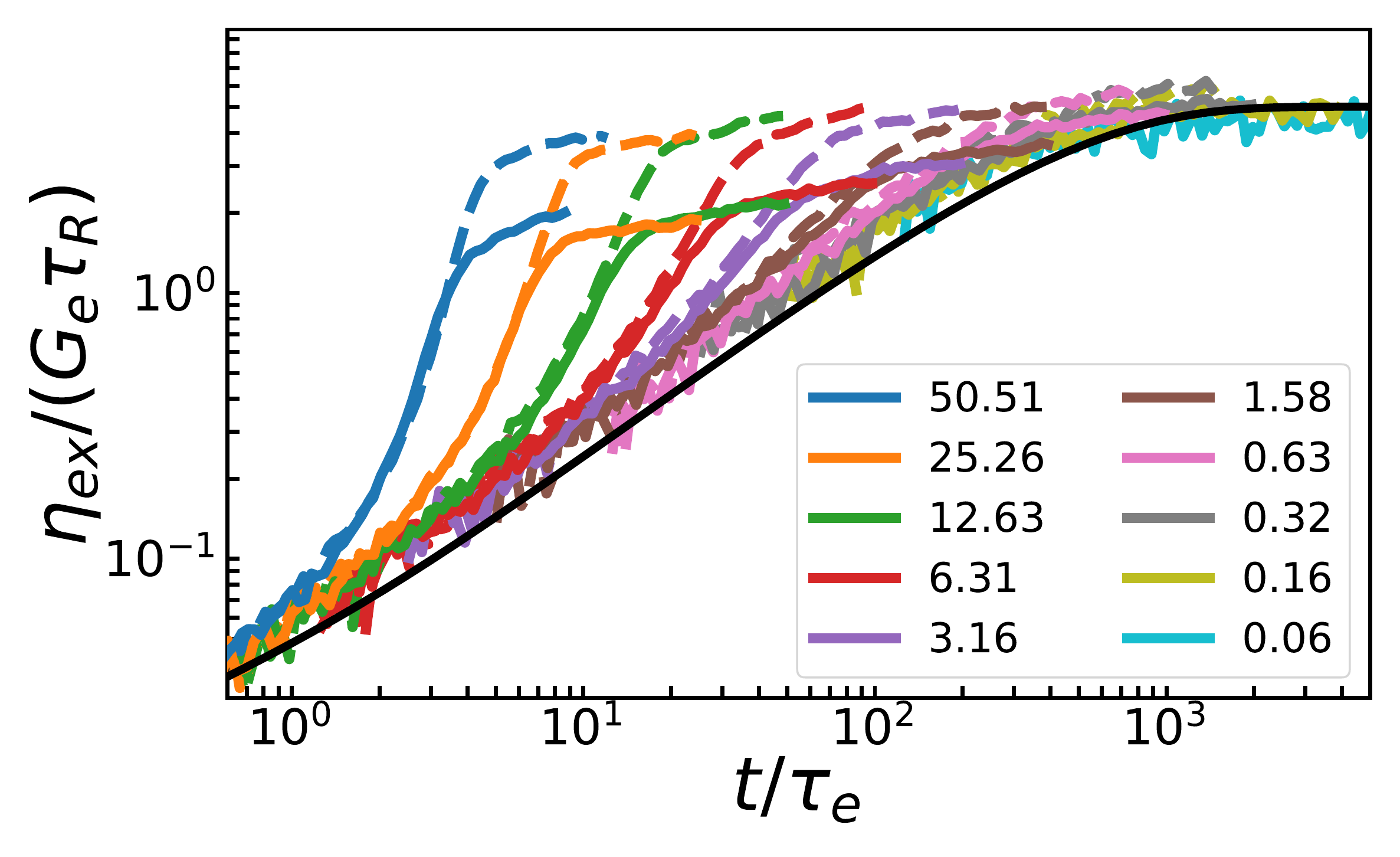}
    \caption{Reduced extensional viscosity $\eta_{ex}$ as a function of time for two melts with $Z\approx 9$ at $Wi_R$ from 0.06 to 52 (colors in legend). The M1 melt has $N=250$, $N_e\approx28$ (solid lines) and the M2 melt has $N=450$, $N_e\approx51$ (dashed lines).
    At low $Wi_R$ both melts approach the analytic prediction from the tube
    model for linear response at $Z=9$ (black line).
    As $Wi_R$ increases, the two melts show increasing deviations from each other and from linear response. The steady-state viscosity at large times, $\bar\eta_{ex}$, shows little change with $Wi_R$ for the M2 melt but drops rapidly with increasing $Wi_R$ for the M1 melt.}
    \label{fig:tdep}
\end{figure}

Figure \ref{fig:tdep} shows how the extensional viscosity $\eta_{ex}\equiv \sigma_{ex}/\dot\epsilon$ evolves with time as polymers are elongated from equilibrium random coils into their steady state conformations at $Wi_R = 0.06$ to $52$.
Results are presented for two melts with the same $Z \approx 9$, but different lengths and entanglement lengths, 
$N_e\approx28$ for M1 and $51$ for M2.
In all cases, the viscosity increases over several decades in time before approaching an asymptotic steady-state value $\bar\eta_{ex}$.
As predicted by tube theory,
the two melts
have nearly the same scaled response at the lowest rates, $Wi_R=0.06$ and $0.16$.
Moreover, the simulations approach the analytic prediction for linear response from the tube model with $Z=9$ (thick black line).

As $Wi_R$ increases, the viscosity grows more rapidly with time than the linear response curve and approaches a new steady-state viscosity at progressively earlier times. The two melts show identical initial deviations from linear response
at each $Wi_R$, but approach different steady state viscosities.
They also exhibit one puzzling aspect of experimental data, the variation of $\bar\eta_{ex}$ with $Wi_R$ is qualitatively different for different polymers \cite{Sridhar2013,Wingstrand2015}.
Here M1 shows a substantial decrease of $\bar\eta_{ex}$ with increasing $Wi_R$, while this shear thinning is nearly absent for $M2$ at this $Z$.

The linear viscoelastic envelope (LVE) shown in Fig. \ref{fig:tdep} is derived with the field theory of Likthman and McLeish for monodisperse, linear polymers \cite{Likhtman2002}. The theory has three parameters: the entanglement time $\tau_e$, the entanglement modulus $G_e=\rho k_b T/N_e$, and the number of entanglements per chain $Z=N/N_e$. We do not fit these parameters to match the analytic theory to our simualtions. Instead, we use the values for $\tau_e$ and $N_e$ measured in previous MD simulation studies \cite{Ge2014,Moreira2015,Hsu2016}. The agreement between simulations and the analytic model supports our use of coarse-grained MD to study entangled polymer rheology. 

\section*{SB. Chain Configurations in Steady State}
\begin{figure*}[h]
    \centering
    \includegraphics[width=\textwidth]{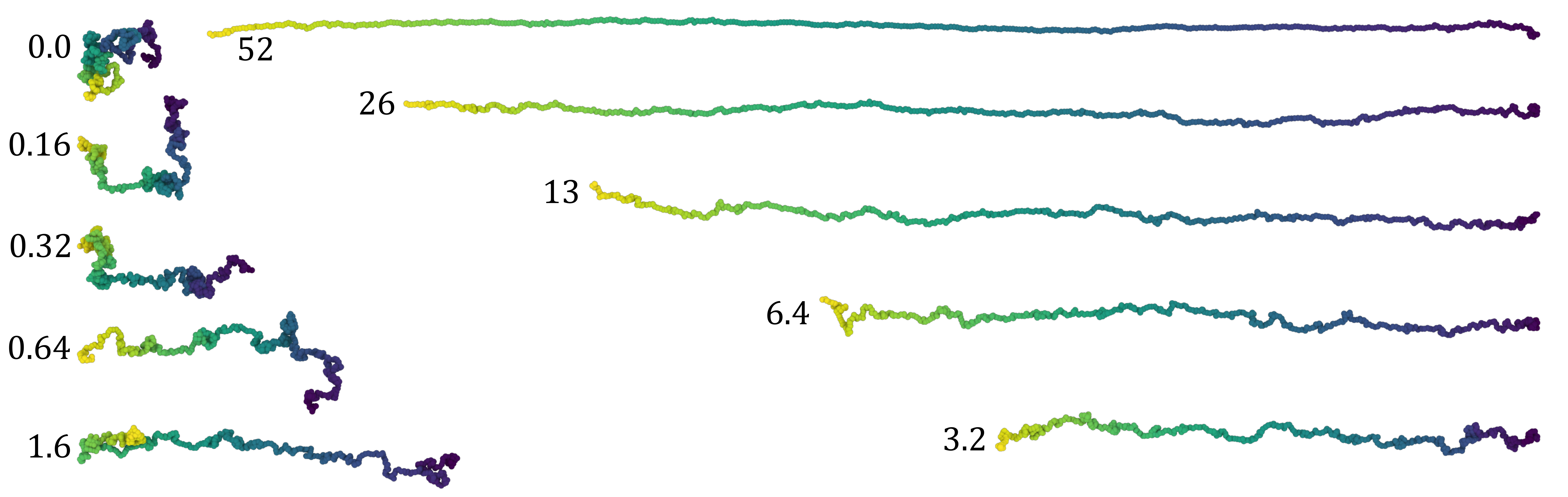}
	\caption{
		 Snapshots of a chain in an M2 melt with $Z=9$ at the indicated $Wi_R$. Note that as $Wi_R$ increases the chains become more aligned along the extension direction and are more confined in the perpendicular plane.
		\label{fig:snapshot}
	}
\end{figure*}

Fig. \ref{fig:snapshot} shows typical chain conformations for an M2 melt with Z=9 in steady-state flows with the indicated $Wi_R$. We select the shown conformations by calculating the steady-state distribution of end-end distances $R(N-1)$ for each $Wi_R$ and selecting a chain with $R(N-1)=\langle R(N-1)\rangle$.

\section*{SC. Steady Orientation and Longest Relaxation Time}
\begin{figure}[h]
    \centering
    \includegraphics[width=0.5\textwidth]{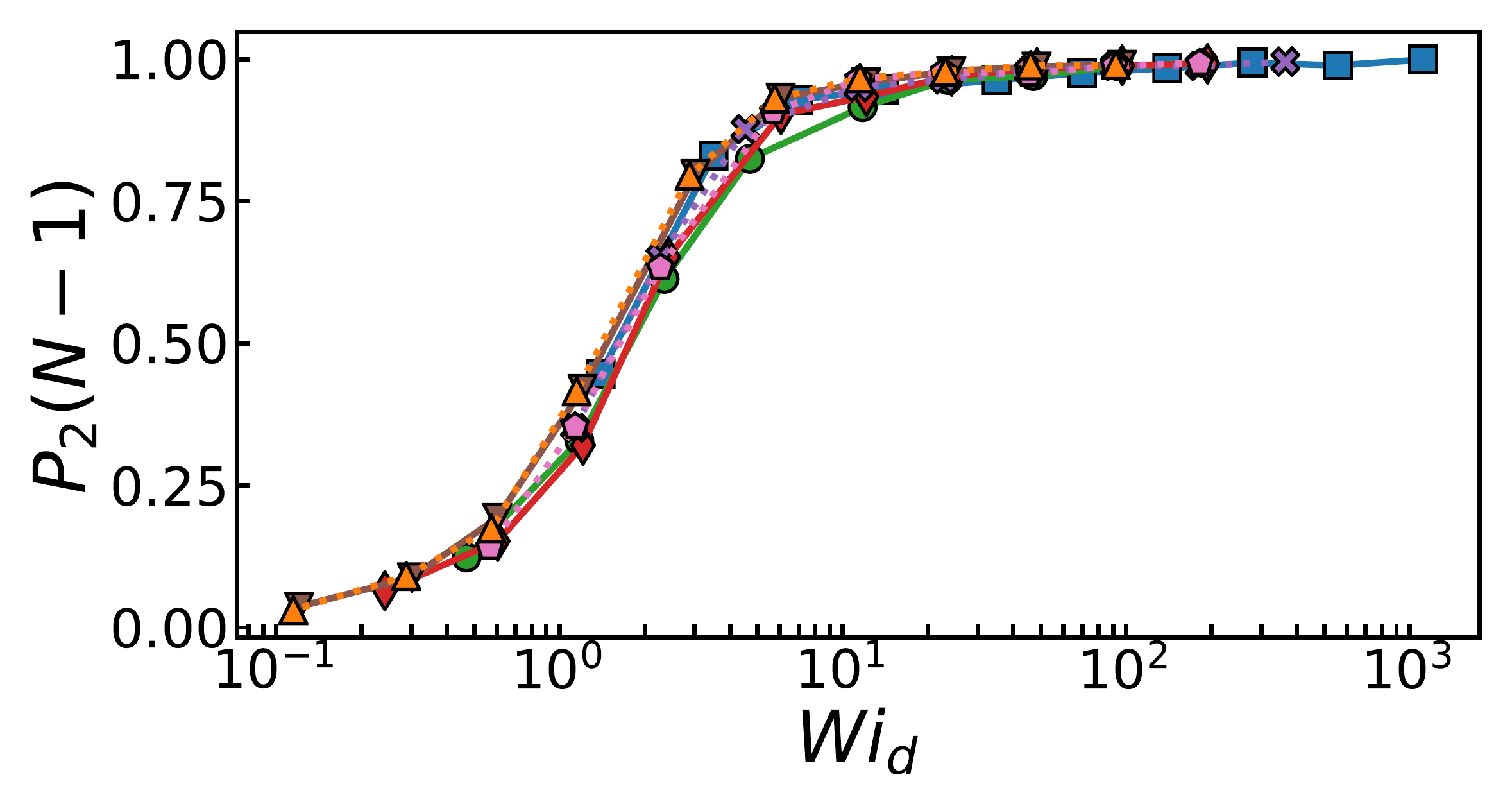}
    \caption{Steady-state end-end vector orientation $P_2(N-1)$ of all melts plotted versus $Wi_d=\dot\epsilon \tau_d$ with $\tau_d$ defined by Eq. \ref{eq:taud}.
    }
    \label{fig:P2Wd}
\end{figure}

The end-end scale orientation of an entangled polymer chain relaxes back to an isotropic equilibrium  $P_2(N-1)=0$ state after the disentanglement time $\tau_d$. 
The disentanglement time is the longest relaxation time for an entangled polymer and is the dominate time-scale entering the Newtonian viscosity.
An expression for $\tau_d$ that includes fluctuations in contour length was derived by Doi and later refined by Likhtman et al. \cite{Doi1983,Likhtman2002}:
\begin{equation}
\tau_d(Z)=3 Z^3 \tau_e 
\left(
1 - \frac{3.38}{\sqrt{Z}} + \frac{4.17}{Z} - \frac{1.55}{Z^{3/2}} + \mathcal{O}(Z^{-2})
\right)
\label{eq:taud}
\end{equation}
Fig. \ref{fig:P2Wd} plots $P_2(N-1)$ versus $Wi_d=\epsilon\dot \tau_d(Z)$ for all melts and $\dot\epsilon$. All data collapse onto a universal curve which shows a rapid increase in orientation at $Wi_d\approx1$. Note, this expression neglects the diffusion of the neighboring chains forming the confining tube. A recent MD study of similar systems found accounting for these effects with the ``double reptation'' approximation accurately described melt viscoelasticity \cite{Hou2010}. This approximation would decrease $\tau_d$ in Eq. \ref{eq:taud} by about a factor of 2 for all melts.

\section*{SD. Measuring Changes in the Tube Radius}

Since the confinement of chains by entanglements is a dynamic many-body effect, a tube radius cannot be uniquely determined from single chain statistics.
Past work on the Kremer-Grest model has used a variety of techniques to analyze the confining tube using primitive path analysis or geometric annealing.
This work provides values of $N_e$, $C_\infty$ and the equilibrium tube diameter
$a_{eq}=R_{eq}(N_e)=C_\infty N_e b^2$, where $R_{eq}(N_e)$ is the rms length of segments of $N_e$ monomers.
The value of $R(N_e)$ 
stops being a useful measure of confinement once chains are extended by flow.
Changes in $R(N_e)$ are dominated by the increase in the length along extension,
while the tube radius is associated with fluctuations about the primitive path
in the perpendicular directions.
There is still debate about how to extend primitive path methods to highly aligned chains, but the tube radius should still scale with perpendicular fluctuations of segments of length $N_e$.  
Measuring these fluctuations becomes simpler at high $Wi_R$ where
the tube is aligned with the extensional axis.
We considered three measures of tube radius that scale with the equilibrium
tube radius at low rates and measure deviations from the extensional axis
at high rates.
As we show here, they all give similar results for the decrease in tube
radius with $Wi_R$.

In Figure 2(d) of the main text we plot a measure of the tube radius based on fluctuations of monomers perpendicular to the end-end vector of segments
of length $N_e$.
The fluctuation $\delta r _ \perp$ depends on the chemical distance $n$ from the start of the segment and must go to zero at both ends.
For Gaussian chains, the maximum fluctuation occurs at $n=N_e/2$ and is equal to $R_{eq}(N_e)/6^{1/2}$.

\begin{figure}[h]
    \centering
    \includegraphics[width=0.5\textwidth]{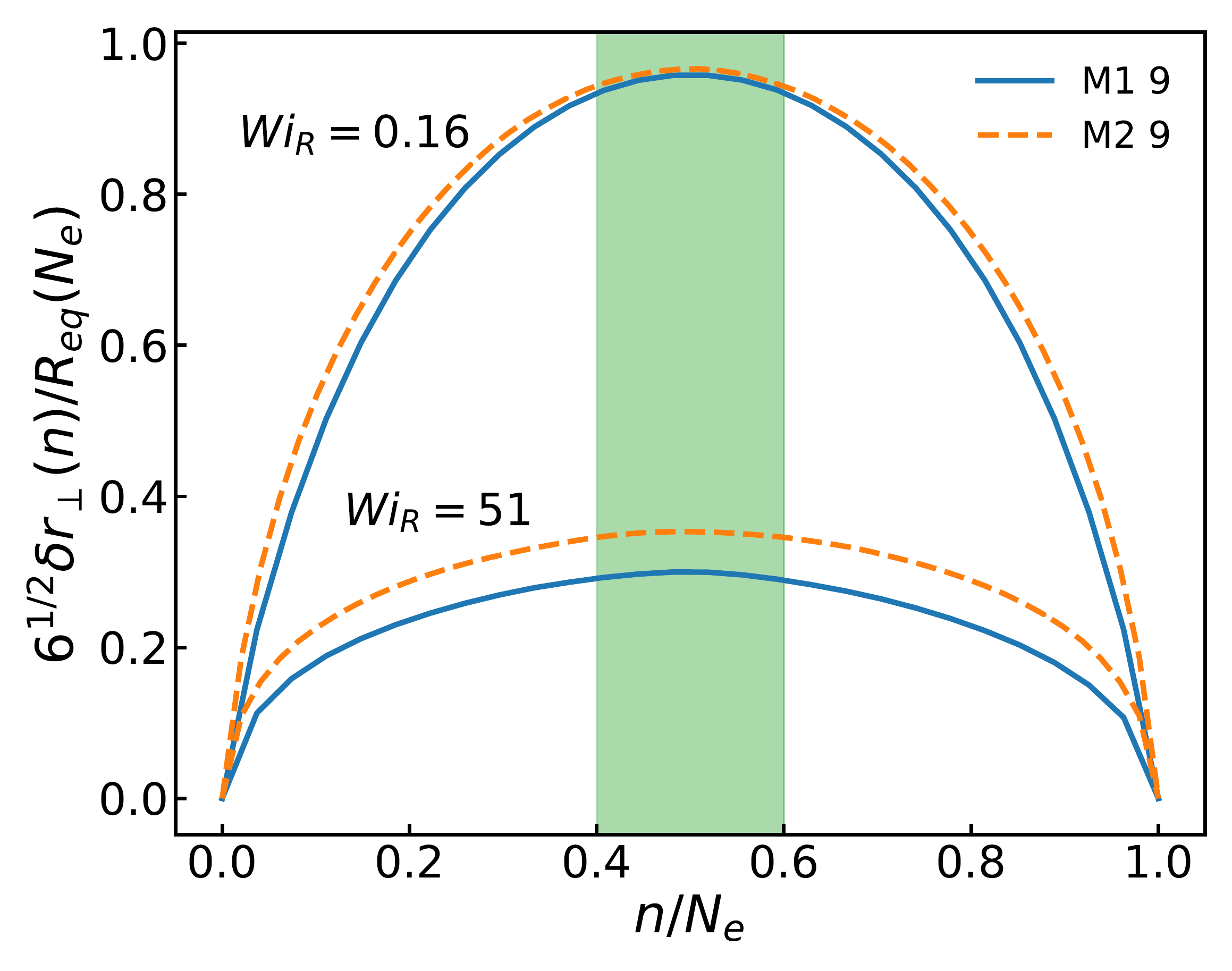}
    \caption{Scaled lateral fluctuation profiles $\delta r_\perp(n)$ at high and low $Wi_R$ for segments of length $N_e$ in M1 and M2 melts. Fluctuations are constrained to go to zero at the two ends and have a central maximum of $R_{eq}(N_e)/\sqrt{6}$ for gaussian coils in equilibrium.
	    Results for $Wi_R=0.16$ are near this equilibrium value.
    The profile decreases and flattens substantially at high rates.
    When computing the maximum lateral fluctuation, we average over the inner 20\% of each segment, indicated by the green shaded region.
    }
    \label{fig:Rperp}
\end{figure}

Average fluctuation profiles $\delta r_\perp(n)$ for each bead in an entanglement segment are shown for M1 (lines) and M2 (dashes) melts at low and high $Wi_R$ in Figure \ref{fig:Rperp}.
Fluctuations are scaled by the equilibrium value at $n/N_e=1/2$ and
$n$ is scaled by $N_e$.
At low rates, the profiles for M1 and M2 are both close to
equilibrium results.
As $Wi_R$ increases, the profiles flatten and $\delta r_\perp$ decreases, signalling a narrowing of the tube.
In Fig. 2(d) of the main text we plot ratios of the average of $\delta r_\perp$
over the middle $20\%$ of each segment (shaded region in Fig. \ref{fig:Rperp}),
corresponding to the inner 6 and 10 beads for M1 and M2, respectively.
This improves our statistics and has negligible effect on the plotted ratios
because the profiles are flat in the center and the
residual curvature produces similar decreases at all $Wi_R$.

\begin{figure}[]
    \centering
    \includegraphics[width=0.75\textwidth]{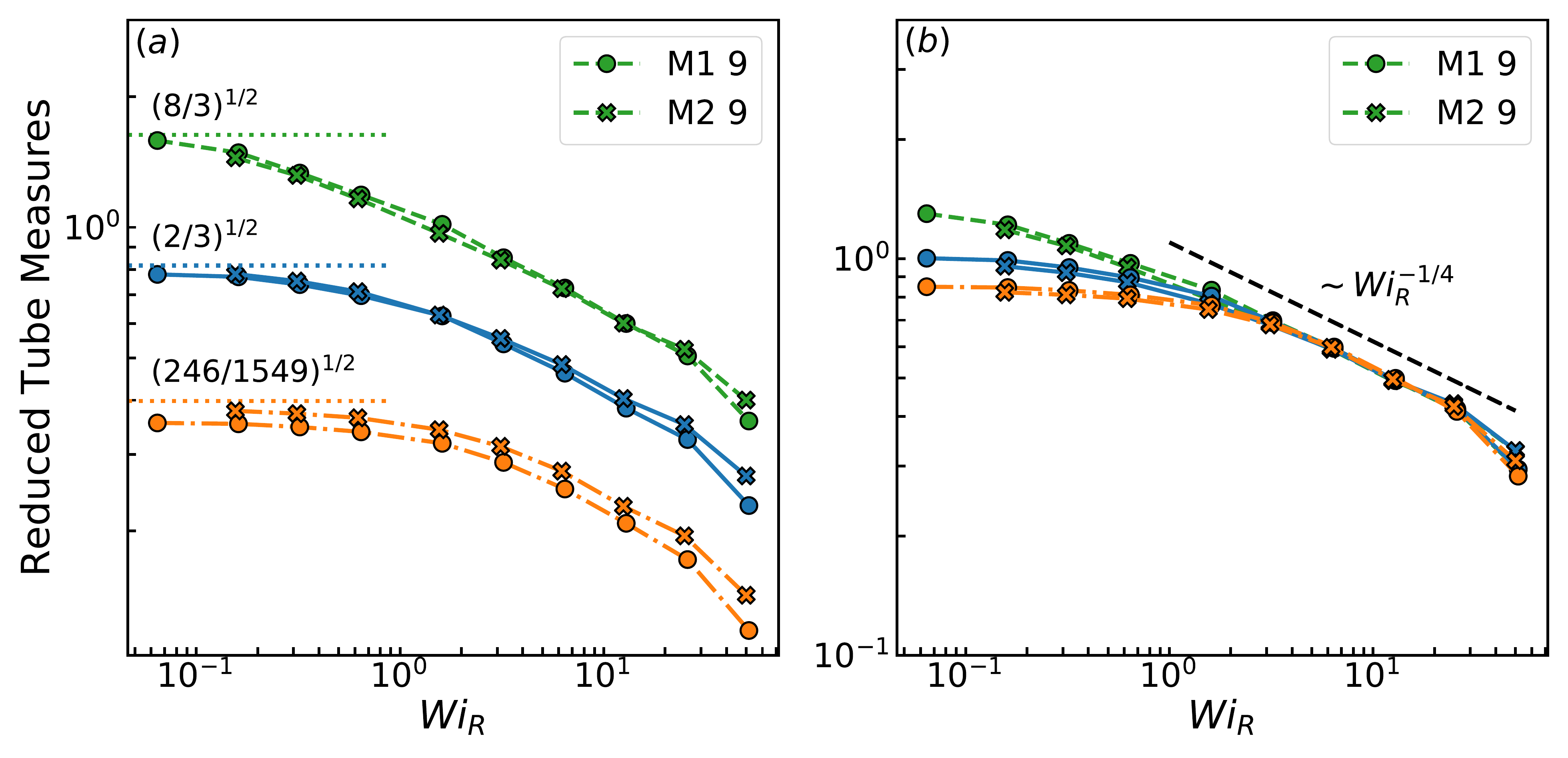}
    \caption{
    (a) Three different chain statistics measuring the lateral tube diameter reduced by the nominal equilibrium value $a_0=R_{eq}(N_e)$ for M1 and M2 melts with $Z=9$.
    (green dashes): The xy-projection of $R(N_e)$ which is perpendicular to the extension axis.
    (blue solid): The maximum RMS deflection $\delta r_\perp$ of segments perpendicular to the segment end-end vector.
    (yellow dash-dots): The average lateral radius of gyration $G_\perp$ computed from the two smallest eigenvalues of each segments radius of gyration tensor.
    Horizontal dotted lines indicate the expected ratio for equilibrium gaussian chains.
    (b) The statistics from (a) collapsed at high $Wi_R$. All statistics exhibit the same rate dependence at large $Wi_R$ and there spread at low $Wi_R$ give bounds for the scaling of the tube diameter at low rate.
    }
    \label{fig:RMeasures}
\end{figure}

Figure \ref{fig:RMeasures} compares the behavior of $\delta r_\perp$ to two
other measures.
One is the length $R_{xy}(N_e)$ of the projection of segments with $N_e$
monomers perpendicular to the extensional axis.
Another is the average minor radius of gyration $G_\perp$ of a segment with $N_e$
beads.
$G_\perp^2$ is the sum of the two smallest eigenvalues of a segment's radius of gyration tensor.
The ratio of the tube diameters from these measures to the equilibrium
tube diameter is readily computed for equilibrium Gaussian chains.
One finds:
$2R_{xy}/a_{eq} = (8/3)^{1/2}$, $2 G_\perp/a_{eq}= (246/1549)^{1/2}$ and 
$2 \delta r_\perp /a_{eq}=(2/3)^{1/2}$.

Fig. \ref{fig:RMeasures}(a) shows that all three measures of tube diameter
approach the equilibrium value at low rates and show similar decreases with
rate for $Wi_R >1$.
We chose to present results for $\delta r_\perp$ in the main text because it is closest to the equilibrium tube radius at low $Wi_R$, is consistent with there
being relatively little change in tube geometry for $Wi_R < 1$ and intermediate between the other measures.

Fig. \ref{fig:RMeasures}(b) shows $\delta r_\perp$ normalized by its equilibrium value as in Fig. 2(d) of the main text.
Also shown are $R_{xy}$ and $G_\perp$.
When scaled by a single constant, all the results collapse for $Wi_R >1$
and show the same power law scaling.
The only difference between the measures is the amount of change at $Wi_R <1$,
and the limiting low rate values differ by less than 20\%.

\section*{SE. Tube Theory Scaling for Steady Extensional Viscosity}

\begin{figure*}[]
    \centering
    \includegraphics[width=0.75\textwidth]{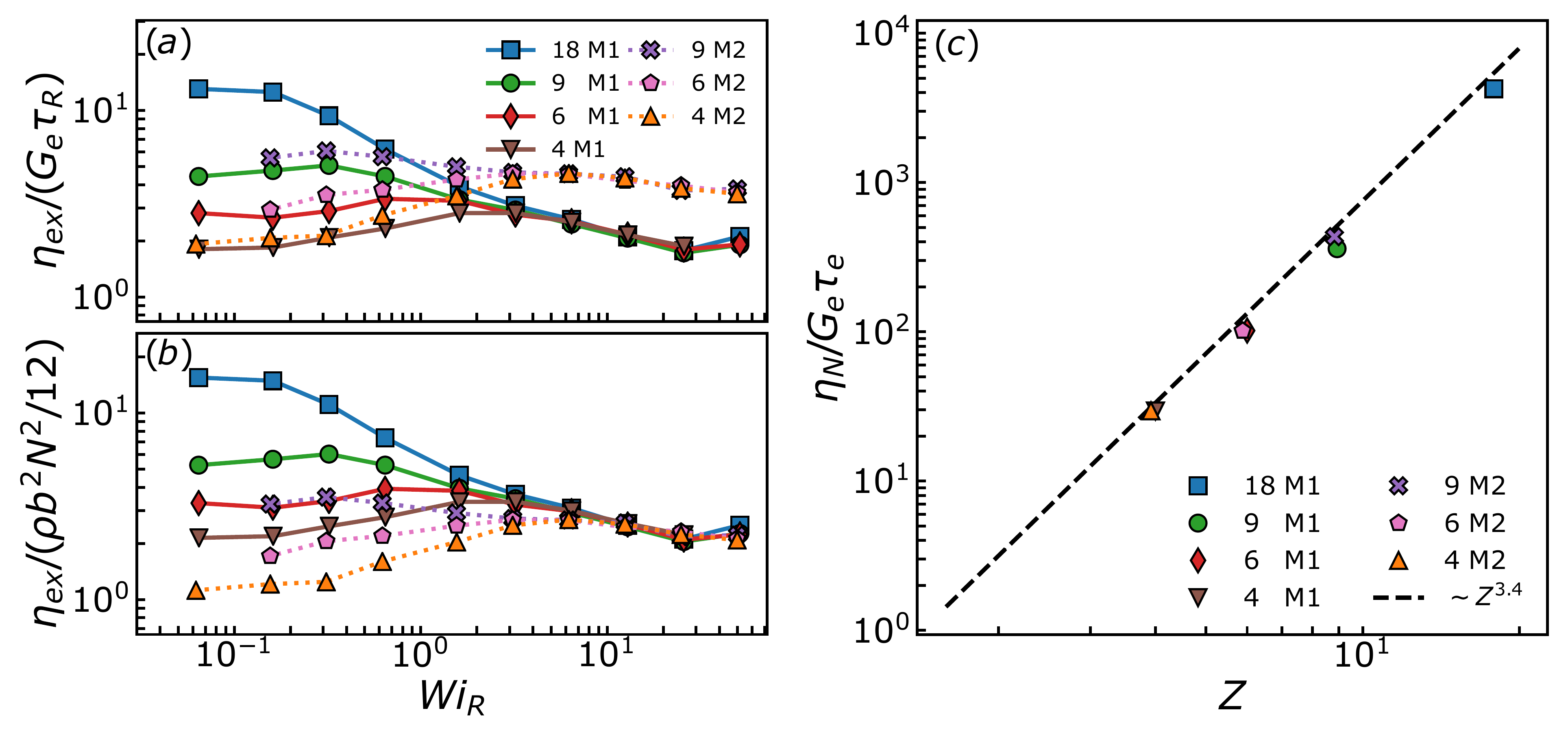}
	\caption{
    (a) Steady state viscosity $\eta_{ex}/G_e \tau_R$ normalized by tube theory parameters.
    (b) Simulation data from panel (a) renormalized by the formula for the asymptotic high rate drag on straight chains. Results for all chains collapse at large $Wi_R$.
    (c) Scaled Newtonian viscosity $\eta_N/G_e\tau_e$ as a function of $Z$. Simulated melts exhibit the approximate $\eta_N\sim Z^{3.4}$ scaling expected for well entangled melts (dashed line).
	\label{fig:repscaling}
	}
\end{figure*}

Reptation theory predicts the scaled viscosity for M1 and M2 melts with the same $Z$ should coincide as $Wi_R\rightarrow0$. 
This is verified by Figure \ref{fig:repscaling}(a), which plots the steady viscosity for all melts in the reduced units of tube theory.
M1 and M2 data with the same $Z$ coincide in the Newtonian limit.
At high $Wi_R$, data for M1 and M2 separate onto two model specific curves, corresponding to the cross over to the high rate scaling discussed in the main text and replotted in Figure \ref{fig:repscaling}(b).

We also verify that $\eta_N$ scales approximately as $Z^{3.4}$ for our systems, as predicted by tube theory for well entangled melts. Figure \ref{fig:repscaling}(c) plots the scaled Newtonian viscosity against $Z$ for all melts. A $Z^{3.4}$ power law is indicated by the black dashed line.

\end{document}